# Infiltration effects on a two-dimensional molecular dynamics model of landslides

Authors: Gianluca Martelloni, Franco Bagnoli

Gianluca Martelloni
Department of Energy Engineering and CSDC (Center of the Study of Complex System)
Via S. Marta 3, 50139 Firenze Italy
Corresponding author: Tel. +39 0554796592; fax: +39 0555609616
E-mail: gianluca.martelloni@unifi.it

Franco Bagnoli
Department of Energy Engineering and CSDC (Center of the Study of Complex System)
Via S. Marta 3, 50139 Firenze Italy
Also INFN, sez. Firenze
E-mail: franco.bagnoli@unifi.it

**Abstract:**
In this paper we propose a two-dimensional (2D) computational model, based on a molecular dynamics (MD) approach, for deep landslides triggered by rainfall. Our model is based on interacting particles or grains and describes the behavior of a fictitious granular material along a slope consisting of a vertical section, i.e. with a wide thickness. The triggering of the landslide is caused by the passing of two conditions: a threshold speed and a condition on the static friction of the particles, the latter based on the Mohr-Coulomb failure criterion (Coulomb 1776; Mohr 1914). The inter-particle interactions are through a potential that, in the absence of suitable experimental data and due to the arbitrariness of the grain dimension is modeled by means of a potential similar to the Lennard-Jones one (Lennard-Jones 1924), i.e., with an attractive and a repulsive part. For the updating of the particle positions we use a MD method which results to be very suitable to simulate this type of systems (Herrmann and Luding 1998). In order to take into account the increasing of the pore pressure due to the rainfall, a filtration model is considered. Finally we also introduce in the model the viscosity as a term in the dynamic equations of motion. The outcome of simulations, from the point of view of statistical and dynamic characterization, is quite satisfactory relative to real landslides behavior and we can claim that this types of modeling can represent a new method to simulate landslides triggered by rainfall.

# 1. Introduction

Landslides are extreme and recurrent events in mountainous areas, often with many implications for urban environments, and consequently on the stricken population, with human casualties and economical losses (Van Asch et al., 2007). Major changes may be induced, sometimes, in a natural environment depending on the extent of the phenomenon. "A major threat is induced by all types of slope movements (e. g., falls, topples, slides, lateral spreads, flows)… which represent one of the most destructive natural hazards on earth" (Brabb 1991). For these reasons landsliding represents a challenging problem in Earth science. Often landslide triggering is caused by an intense and/or long rain. In particular, shallow landslides are triggered by short intense rainfalls (Campbell 1975; Crosta and Frattini 2007), while deep landslides are connected with prolonged and less intense rainfall events (Bonnard and Noverraz 2001). Thanks to the rapid development of computers and advanced numerical methods, physical based models have been developed to predict the landslide triggering and to evaluate the run-out. Two fundamental approaches have been proposed to assess the dependence of landslide triggering on rainfall measurements. The first one relies on deterministic models (infiltration and geotechnical based) while the second defines the statistical rainfall thresholds above which the triggering of one or more landslides is possible (Segoni et al., 2009; Martelloni et al., 2011; Rosi et al., 2012). Regarding the propagation of a landslide, most of the numerical methods have used a continuum approach, i.e., an Eulerian point of view (Crosta et al. 2003, Patra et al. 2005). Other modeling approaches are based on cellular automata (Avolio et al., 2008). A relatively less common approach is the Lagrangian one, based on discrete-particle methods, in which the material forming the slope (and the landslide) is represented as an ensemble of interacting elements, called particles or grains. The discrete element method (DEM) is used to model granular materials, debris flow and flow-like landslides (Cundall and Strack (1979); Iordanoff et al. 2010). Another Lagrangian method is the molecular dynamics (MD) one, closely related to DEM. This latter method is generally distinguished by the inclusion of rotational degrees-of-freedom as well as stateful contact and often complicated geometries. The inclusion of a more detailed description of the elementary components or their interactions and, above all, the increasing of the number of elements of the system allows for more realistic simulations, but the computational load can be very onerous. Obviously, the accuracy of the simulation has to be compared with the available experimental data. In the case of laboratory experiments, very accurate data can be obtained, but this is not possible for real landslides. These arguments motivated us to reduce the complexity of the model as much as possible, examining if this choice is compatible with the behavior of real landslides. In  previous works we proposed a shallow landslide modeling (Massaro et al., 2011; Martelloni et al., 2012). In this paper we present an integration of a filtration model into a MD model for the starting and prosecution of particle movement along a slope, after a triggering induced by rainfalls. This model is conceived to be characteristic of deep landslides. The inclusion of the rainfall effect, i.e., the modeling of the effects due to the fluid that filters in the porous of the material, causing the landslide triggering, is a challenging problem. Our idea is to integrate the Iverson model of infiltration (Iverson, 2000) with the MD approach, by considering the infiltration at the particle level where we use a failure criterion of Mohr-Coulomb to assess the local triggering within the slope. Moreover we introduce in the model some stochastic variations to take into account the variability of the slope in terms of the water infiltration and frictional behavior. At present we do not pretend to be able to develop a model that simulates a real landslide or debris flow, rather we want to explore new alternative approache useful for this kind of problems. The resulting numerical method, similar to that of molecular dynamics (MD), is based on the use of an interaction potential between the particles, similar to the Lennard-Jones one. As we shall see in the following sections, by means of this type of force we can also simulate a compressed state of the particles, according to a stress state of the slope material.

# 2. Modeling approach

## 2.1 Filtration modeling and triggering mechanism

In a previous work (Martelloni et al., 2012) we proposed a model for shallow landslides triggered by rainfall. This model is coarse-grained, based on fictitious particles, using a molecular dynamic approach for the update. In this previous version we considered only one particle layer. Due to this reason and to the quick response to rainfall of shallow landslide, we did not introduce there an infiltration model to integrate the triggering dynamics, although also for shallow landslides the triggering mechanism is related to pore pressure increasing. Obviously, in case of deep landslide this choice cannot be made and therefore we extended the model by including the crucial role of increasing pore pressure due to the rain infiltration, that is the main actor of the triggering mechanism (van Asch et al., 1999). At present we use the Iverson filtration model (Iverson 2000) that is adapted to the molecular dynamics approach according to the failure criterion of Mohr-Coulomb.

The idea is to use the one-dimensional infiltration equation along the $z$ coordinate of the reference system ($x$-$z$) along the slope (Fig. 1):

$$\frac{\partial \mu(z,t)}{\partial t} = K \cdot \frac{\partial^2 \mu(z,t)}{\partial z^2} \qquad (1)$$

where $\mu(z,t)$ is the pore pressure at depth $z$ (in Eq. 1 the $z$ coordinate is reverse with respect to Fig. 1) and time $t$, while $K$ is the diffusion coefficient depending on slope angle $\alpha$ that is held a constant in our simulations.

At the time $t$ = 0 the particles are arranged on a regular grid and the material is initially dry, i.e., it exhibits an initial pore pressure distribution equal to zero. Starting from time $t$ = 0, a constant rain is simulated, but for each vertical layer $n_j$ (Fig. 1) we assume a different infiltration (small stochastic variations) along $x$ axes of the slope

(Fig. 2). According to Eq. 1, the solution is given by the rainfall input per response function, i.e.,

$$\begin{cases} \mu(z,t^*) = \dfrac{I_z}{K_z} \cdot R^* \\ R^* = \begin{cases} R(t^*), & t^* \le T^* \\ R(t^*) - R(t^* - T^*), & t^* > T^* \end{cases} \\ R(t^*) = \sqrt{t^*/\pi} \cdot \exp(-1/t^*) - erfc(1/\sqrt{t^*}) \end{cases} \tag{2}$$

where $t^*$ and $T^*$ are respectively the normalized time and the normalized rainfall duration (Iverson 2000), while $I_z$ and $K_z$ are respectively the average infiltration rate and the hydraulic conductivity in the slope-normal direction.

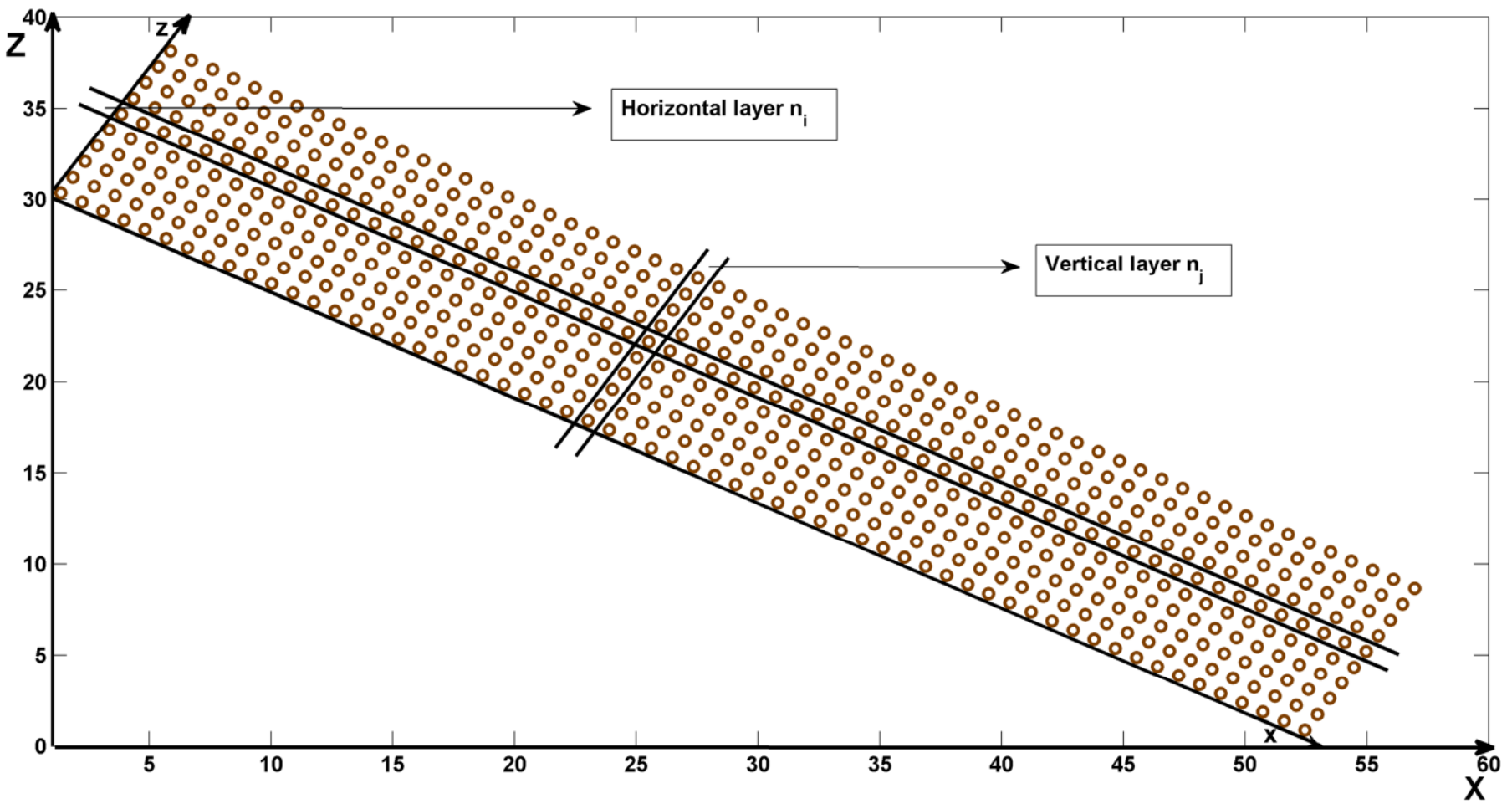

**Figure 1** Reference system (*x*-*z*) of the slope modeled with particles arranged in a regular grid according to disposition in horizontal and vertical layer

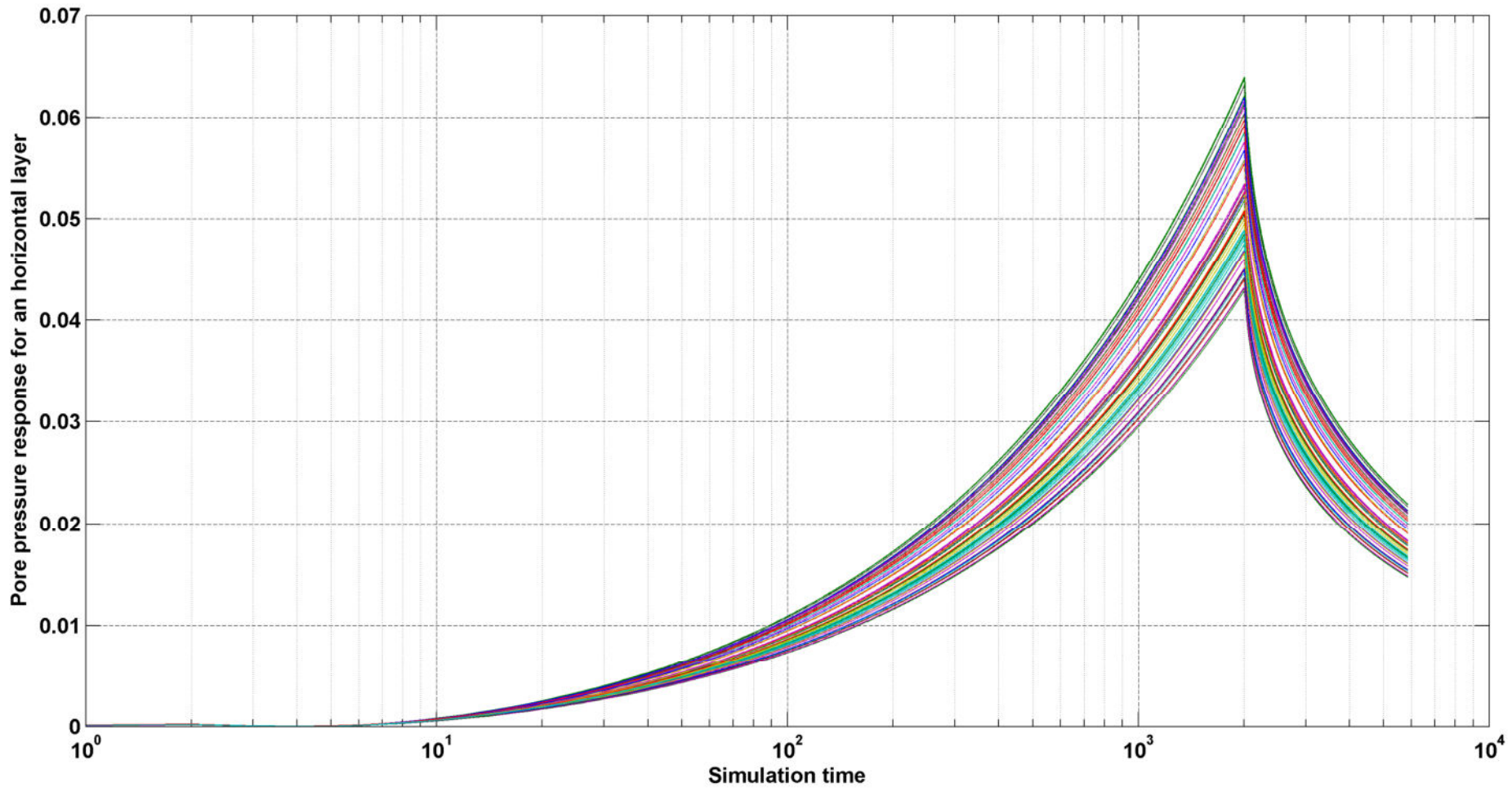

**Figure 2** The pore pressure response, in a simulation of our system, for each position x of a horizontal layer ni in the time steps of simulation: the differences are due to stochastic variations

Now let us show how the infiltration model is integrated into our numerical scheme based on molecular dynamics (a discrete Lagrangian approach similar to DEM). The first step is to appropriately relate the simulated rainfall with the water content of the particles that constitute our fictitious soil. Let $I_z$ be the discrete infiltration rate, i.e.,

$$I_z = K_z \cdot \frac{\Delta h}{\Delta z} \tag{3}$$

We assume that $\Delta h$ is the rainfall increment at a generic instant $t$ and thickness $z$, while $\Delta z$ is the initial distance between the centers of mass of two adjoining particles. It is now possible to deduce $I_z$ in terms of a mass ratio from simple considerations on the density that can be expressed as:

$$\rho = \frac{dm}{dV} = \frac{dm}{S \cdot dh} \tag{4}$$

where $S$ is the unit area, $m$ the mass and $V$ the volume. Let us consider the density of water $\rho_w$ and the density of fictitious material (i.e. the particles) $\rho_p$. We obtain the infinitesimal heights $dh_w$ and $dh_p$,

$$\begin{cases} \rho_w = \dfrac{dm_w}{S_w \cdot dh_w} & \Rightarrow \quad dh_w = \dfrac{dm_w}{S_w \cdot \rho_w} \\ \rho_p = \dfrac{dm_p}{S_p \cdot dh_p} & \Rightarrow \quad dh_p = \dfrac{dm_p}{S_p \cdot \rho_p} \end{cases} \tag{5}$$

but, in our case, since the dimensions of particles are relatively small, it is possible to consider, to a good approximation,

$$S_w \cdot \rho_w = S_p \cdot \rho_p \tag{6}$$

Consequently, the ratio between $dh_w$ and $dh_p$ gives the new discrete infiltration rate $I_z$:

$$\left\{ \frac{dh_w}{dh_p} = \frac{dm_w}{dm_p} \right. \quad \Rightarrow \quad \frac{\Delta h_w}{\Delta h_p} = \frac{\Delta m_w}{\Delta m_p} \quad \Rightarrow \quad I_z = K_z \cdot \frac{\Delta m_w}{\Delta m_p} \tag{7}$$

Hence, we can simulate the rainfall in terms of water mass and, using the response function $R^*$, we can take into account the absorbed water in time and space at thickness $z$, i.e., at each level of the particle layers.

Consequently, since the gravity acts on each particle $i$, its components can be expressed along the slope reference system as:

$$\mathbf{F_{gi}} = \{ g \cdot \sin(\alpha) \cdot [m_i + w_i(t)], \quad -g \cdot \cos(\alpha) \cdot [m_i + w_i(t)] \} \tag{8}$$

where $g$ is the gravity acceleration, $\alpha$ the angle of the slope, $m_i$ is the dry mass, variable from particle to particle and $w_i(t)$ is the cumulative absorbed water in time. The interaction force $\mathbf{F_{ij}}$, that acts on particle $i$ due to particle $j$, is defined trough a potential inspired to the Lennard-Jones one, i.e., we consider that the repulsive and attractive term of the potential or force are weighted differently:

$$\mathbf{F_{ij}} = -\mathbf{F_{ij}} = -\left[k_1\left(\frac{r}{R_{ij}}\right)^{-2} - k_2\left(\frac{r}{R_{ij}}\right)^{-1}\right]\cdot\hat{\mathbf{r}}$$
$$R_{ij} = L = 1 \qquad (9)$$
$$r = \left|\mathbf{r_{ij}}\right| = \sqrt{\left(x_j - x_i\right)^2 + \left(y_j - y_i\right)^2}$$

where $r$ is the distance between the centers of mass, $k_1$ and $k_2$ are constants ($k_1 = k_2$ in "classical" Lennard-Jones potential), $\hat{\mathbf{r}}$ is the unit vector relative to the force and $L$ is the equilibrium distance (Fig. 3). If $k_1 = k_2$ we have the equilibrium at distance $L = 1$ (Fig. 4), else if $k_1 \neq k_2$ it is possible simulate, starting from $t = 0$, a compressed stress state of the particles (Fig. 5). The justification of such interaction force is due simulation results that are similar to real landslide behavior (see simulations section). As mentioned previously, at instant $t = 0$ the system is prepared in equilibrium, that is, the particles are disposed on a regular grid (Fig. 1). Therefore, as triggering mechanism, we consider the law of Mohr-Coulomb (Coulomb 1776; Mohr 1914) in the form of the effective stress (Terzaghi 1943),

$$\tau_f = (\sigma - \mu)\cdot\tan\phi' + c' \qquad (10)$$

where $\tau_\phi$ is the shear stress at failure, $\sigma$ the normal stress, $\phi'$ the friction angle and $c'$ the cohesion term. As the Mohr-Coulomb failure criterion is a simple friction law, short of the term of cohesion, it can be easily adapted to our case, rewriting Eq. (10) as follows:

$$\tau_f = F_s + c' = \left[M(z,t)\cdot g\cdot\cos(\alpha) - \mu(z,t)\right]\cdot\mu_s + c' \qquad (11)$$

where $M$ is the term $m_i$+$w_i(t)$, relative to particle $i$, considered in Eq. (8). Note that varying the thickness $z$, the considered particle layer differs due to the discreteness of the system.

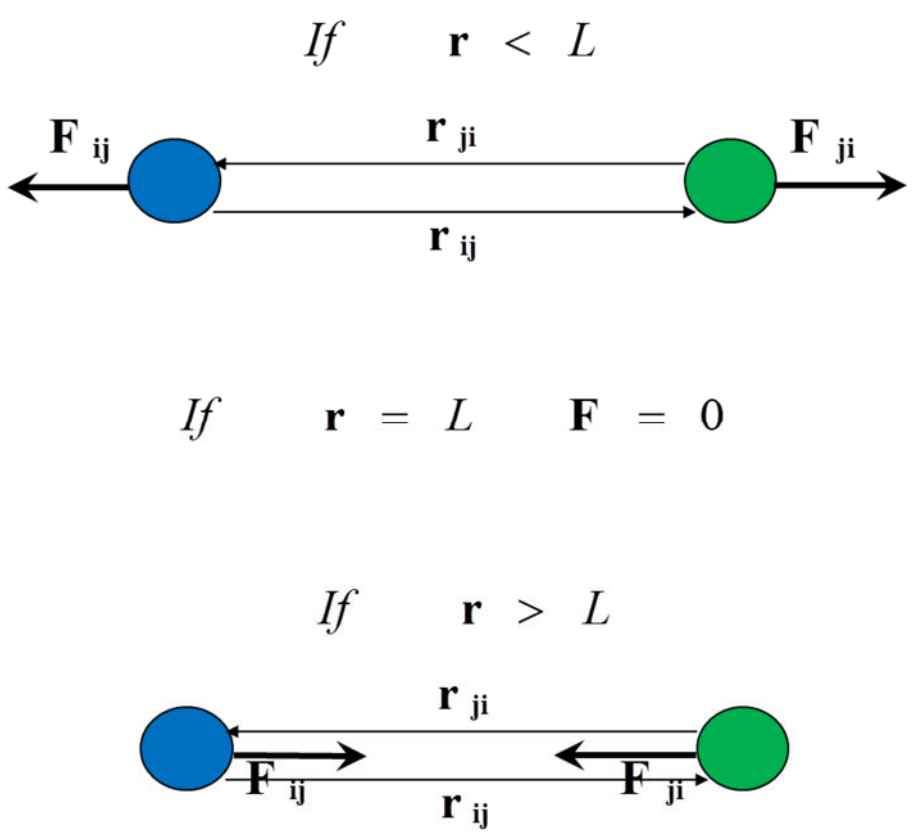

**Figure 3** Schematic description of the interaction force

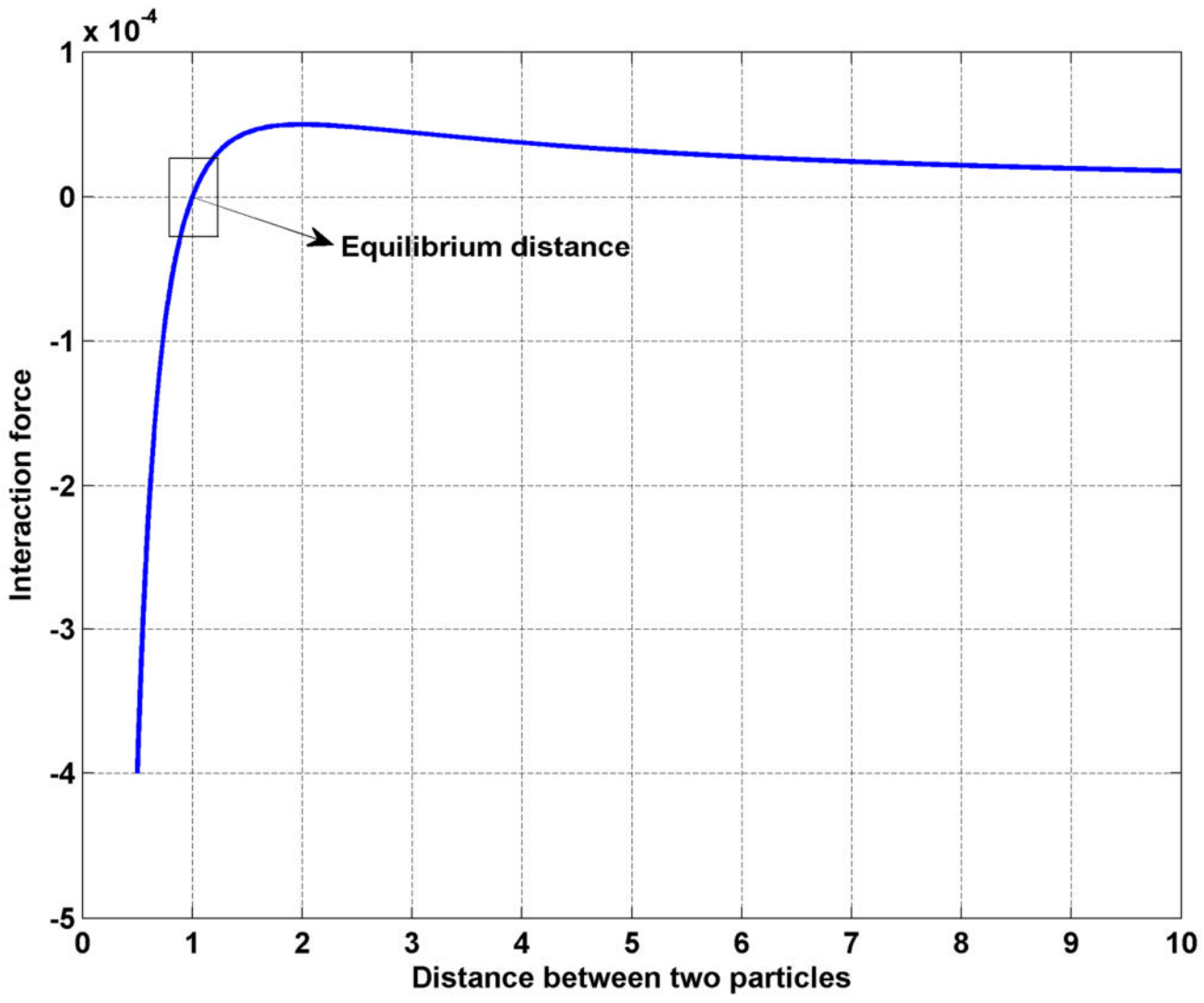

**Figure 4** Interaction force for the equilibrium distance $L = 1$

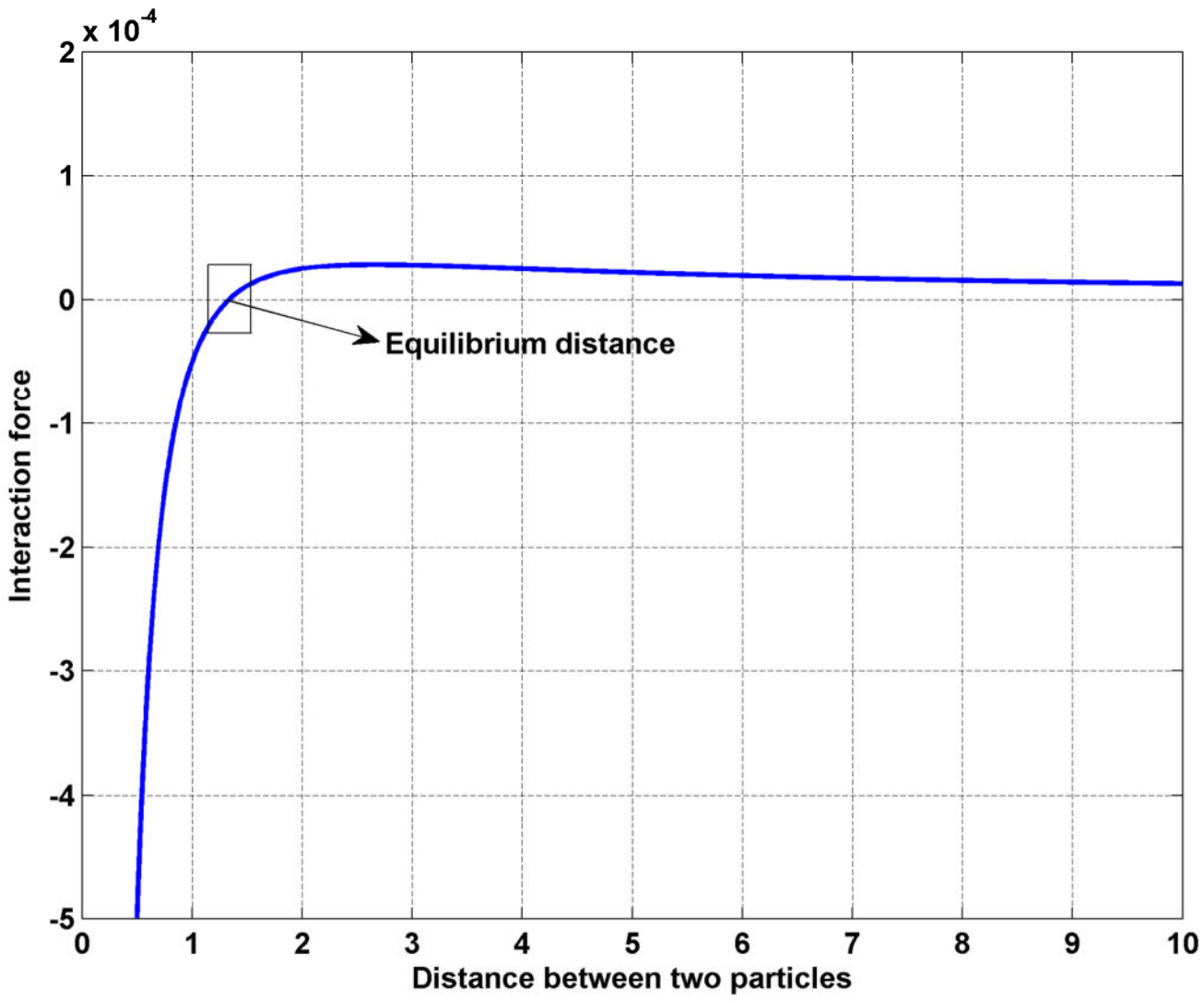

**Figure 5** Interaction force for the equilibrium distance $L > 1$ (simulated initial compressed stress state)

Finally, in our model, we express the local triggering, i.e., the triggering at the particle level, using a failure criterion and considering a speed threshold $v_d$ for the static-dynamic transition. In synthesis, for each particle $i$:

$$\begin{cases} |\mathbf{F_i}| < F_{si} + c_i' \\ |\mathbf{v_i}| < v_d \end{cases} \tag{12}$$

$$\mathbf{F_i} = \mathbf{F_{gi}} + \sum_{\mathbf{j}=1}^{\mathbf{j}=n_k} \mathbf{F_{ij}} \tag{13}$$

where $|\mathbf{F_i}|$ represents the module of the active forces, i.e., the force of gravity $\mathbf{F_{gi}}$ plus the force resulting from the potential, the latter being the sum of terms in Eq. (13) where $n_k$ denote the total number of particles among the next-to-nearest neighbors in interaction with particle $i$, i.e., for initial instant, $n_k = 8$ (Fig. 6), while $|\mathbf{v_i}|$ is the module of the speed. This double control, expressed by Eqs. (12) on the forces and velocity, permits both the triggering and the stopping of the particle motion.

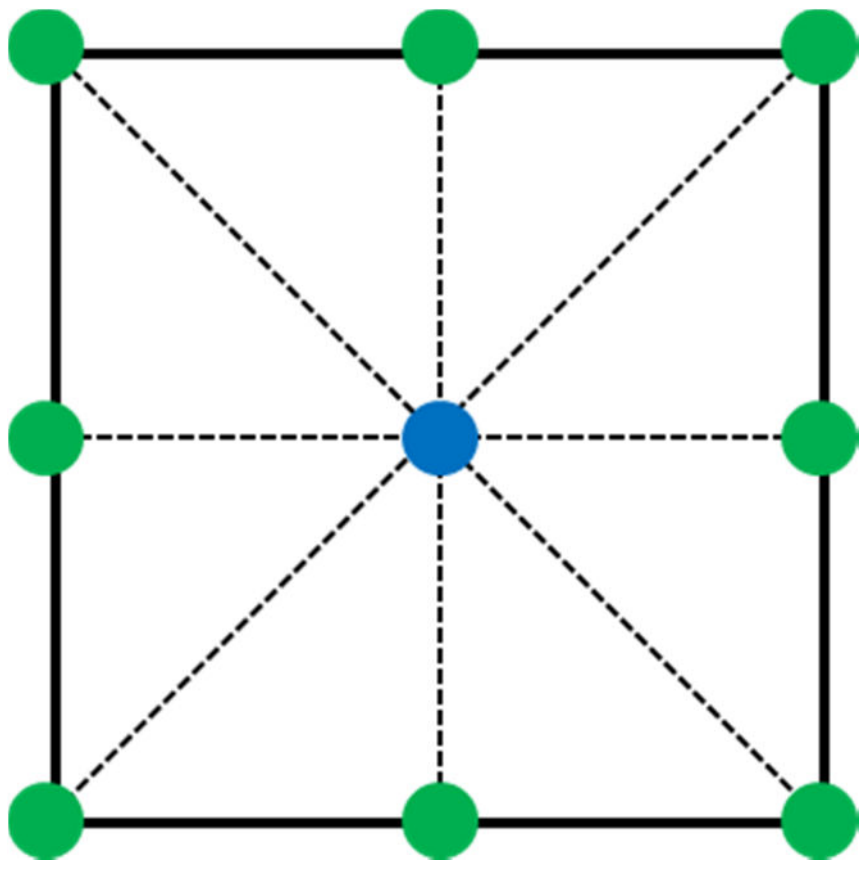

**Figure 6** Interaction, at instant $t = 0$ for each particle of the system, according to second neighbors one

## 2.2 Dynamics condition and updating algorithm

Eqs. (12) are valid in dynamical conditions, as they represent, in synthesis, a control on the state of motion of the particles. Once a particle is moving, we consider also, as active force, a dynamic friction (the force direction is opposed to the velocity one), expressed for each particle $i$ by:

$$\mathbf{F_{di}} = (m_i + w_i(t))g\cos(\alpha)\cdot(\mu_d \cdot \exp(-w_0 t) + \mu_{dlow} \cdot (1-\exp(-w_0 t)))\cdot(-\hat{\mathbf{v}}) \tag{14}$$

The force in Eq. (14) depends on two friction terms, characterized by coefficients $\mu_d$ and $\mu_{dlow}$, i.e. $\mu_d$ for $t = 0$ and $\mu_{dlow}$ for $t \rightarrow \infty$, with $\mu_d > \mu_{dlow}$. In synthesis, the effect of rainfall is to decrease the friction of the particles during time (through the constant velocity $w_0$ of the exponential). Moreover the friction coefficients $\mu_d$ and $\mu_{dlow}$ vary randomly (with a small dispersion) with the position, modeling the roughness between the particles. This friction law is inspired by Jop et al. (2006).

As previously mentioned, initially the particles are arranged on a regular grid, i.e., at the instant $t = 0$ each mass is placed in the nodes of a regular rectangular grid and therefore every particle interacts with the eight blocks placed in the nearest and next-to-nearest nodes (Fig. 6). At each time step, the interactions are re-calculated for each object within a given interaction range. This technique is used in molecular dynamics and congruent with principle of action and reaction (Fig. 7).

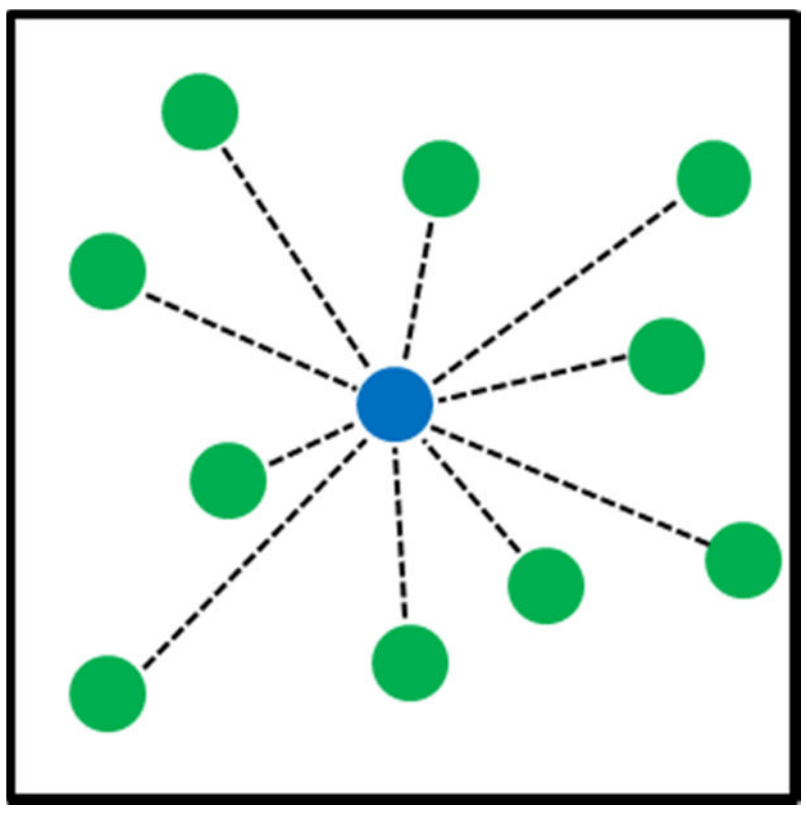

**Figure 7** Ri-calculus of interaction for each mass within the assigned range

Generally, in MD and accordingly in our simulations, the updating of the positions and velocities is based on the first or second-order Verlet algorithm (Verlet 1967). The latter is very stable, allowing a good numerical approximation. Moreover, as the forces are calculated once for each time step, this computational updating method does not require a large computational power.

When a mass is moving, the total force **F** that acts on it is given by the sum of the active forces, the dynamic friction force and a term of viscosity with coefficient $\mu$,

$$\mathbf{F} = \mathbf{F_i} + \mathbf{F_{di}} + \mathbf{F_{\gamma i}} = \mathbf{F_{gi}} + \sum_{\mathbf{j}=1}^{\mathbf{j}=n_k} \mathbf{F_{ij}} + \mathbf{F_{di}} - \mu \cdot \mathbf{v_i} \qquad (15)$$

In this case, the value $n_k$ in the sum of Eq. (15) can be less or grather than 8, due to the possible compression effects during the motion of masses.

The velocity Verlet algorithm, for the updating of positions **r** and velocities **v** of each particles between two instant of difference $\Delta t$, reads

$$\begin{cases} \mathbf{r}(t+\Delta t) = \mathbf{r}(t) + \mathbf{v}(t)\Delta t + \frac{\mathbf{F}(t)}{m}\Delta t^2 \\ \mathbf{v}(t+\Delta t) = \mathbf{v}(t) + \frac{1}{2m}\left[\mathbf{F}(t+\Delta t) + \mathbf{F}(t)\right]\Delta t \end{cases} \quad (16)$$

Summing up we note that, in the case of uniform rainfall, it is simple to theoretically deduce the time of local triggering, i.e., the time of the first particle detachment. However, since the sliding masses could stop after a first detachment, the triggering of single particle cannot represent the definition of landslide triggering. A better definition in this sense is based on the motion of center of mass of the global system or the center of mass of all particle in motion (Martelloni et al. 2012). In the next section we see that is possible to use a Fukuzono method (Fukuzono 1985) to predict the failure time for our simulated system.

# 3. Results of model simulations

In this section we show the simulation results, and exhibit some peculiarities that emerge from the analysis of generated data. Regarding the dynamics, we observe the typical stick-and-slip dynamics of frictional systems, earthquake faults and landslides (Nielsen et al., 2010) that is also observed in other MD model as the seismic fault one (Ciamarra et al., 2010). In Figs. 8 and 9 the mean kinetics increment of the particles and the mean velocity are reported, respectively. It is possible to note a first stick phase and a subsequent slip one (Heslot et al., 1994). In Fig. 10, the time behavior of the inverse of the mean velocity is plotted and there we can note better the initial stick phase. The behavior of this simulation, in terms of the velocity, is similar to real landslide (Suwa et al., 2010). As mentioned above, we use the Fukozono method of the inverse of velocity for the evaluation of failure time of simulated landslide. Let us apply first this method to the initial part of the simulation, corresponding to the maximum slope of the inverse of velocity (green circle in Fig. 11) up to consider all the points (red circle in Fig. 11), evaluating the time of triggering by means of the calibration function,

$$\frac{1}{v} = [\beta \cdot (\alpha - 1)]^{\frac{1}{\alpha-1}} \cdot (t_r - t)^{\frac{1}{\alpha-1}} \tag{17}$$

where $v$ is the mean velocity of the simulated landslide (i.e., the masses in motion), $t$ the time of simulation, $t_r$ the time of failure, while $\alpha$ and $\beta$ are constant. These evaluated triggering times vary from 150 to 220 simulation time steps.

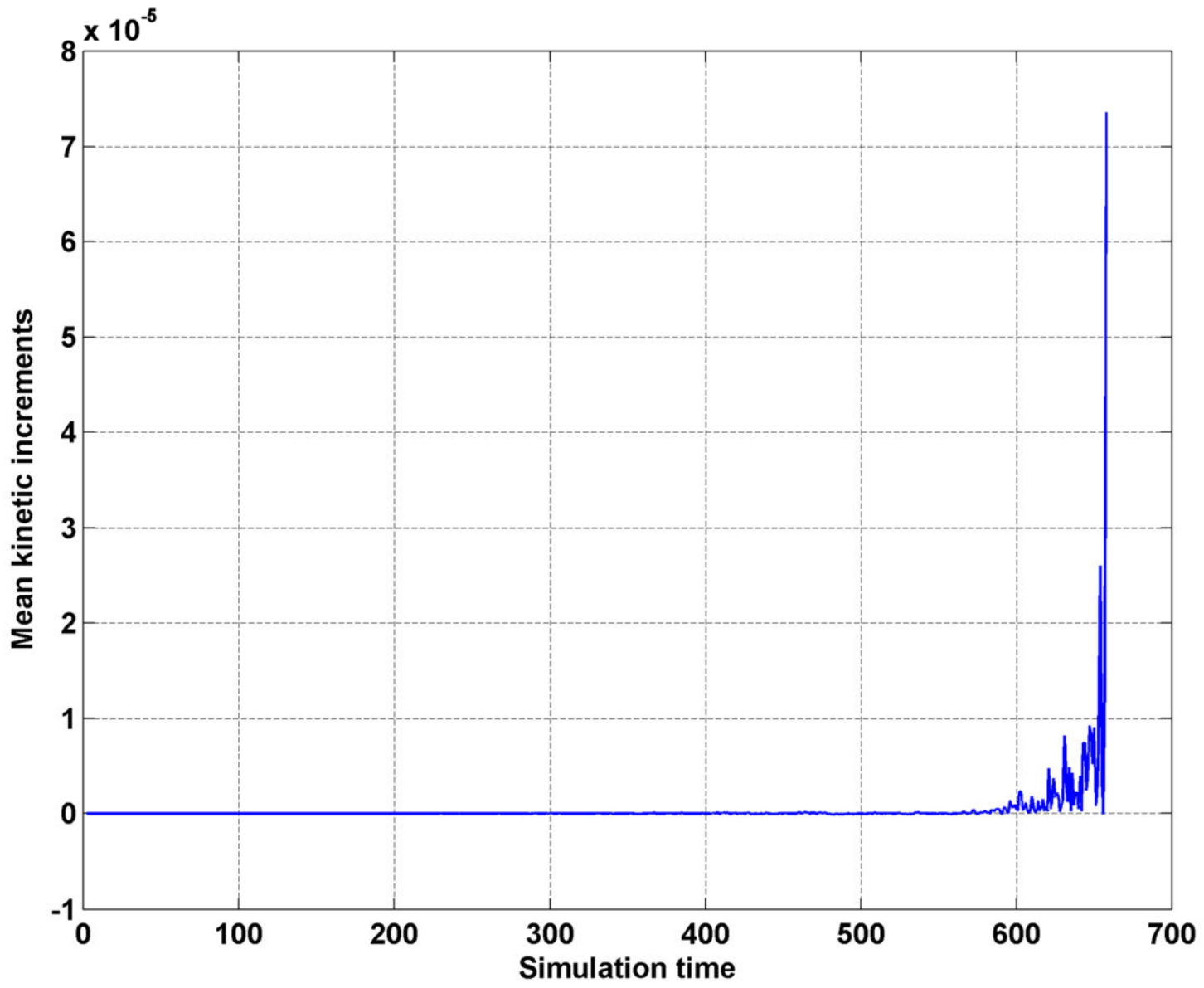

**Figure 8** Mean kinetic increment versus simulation time

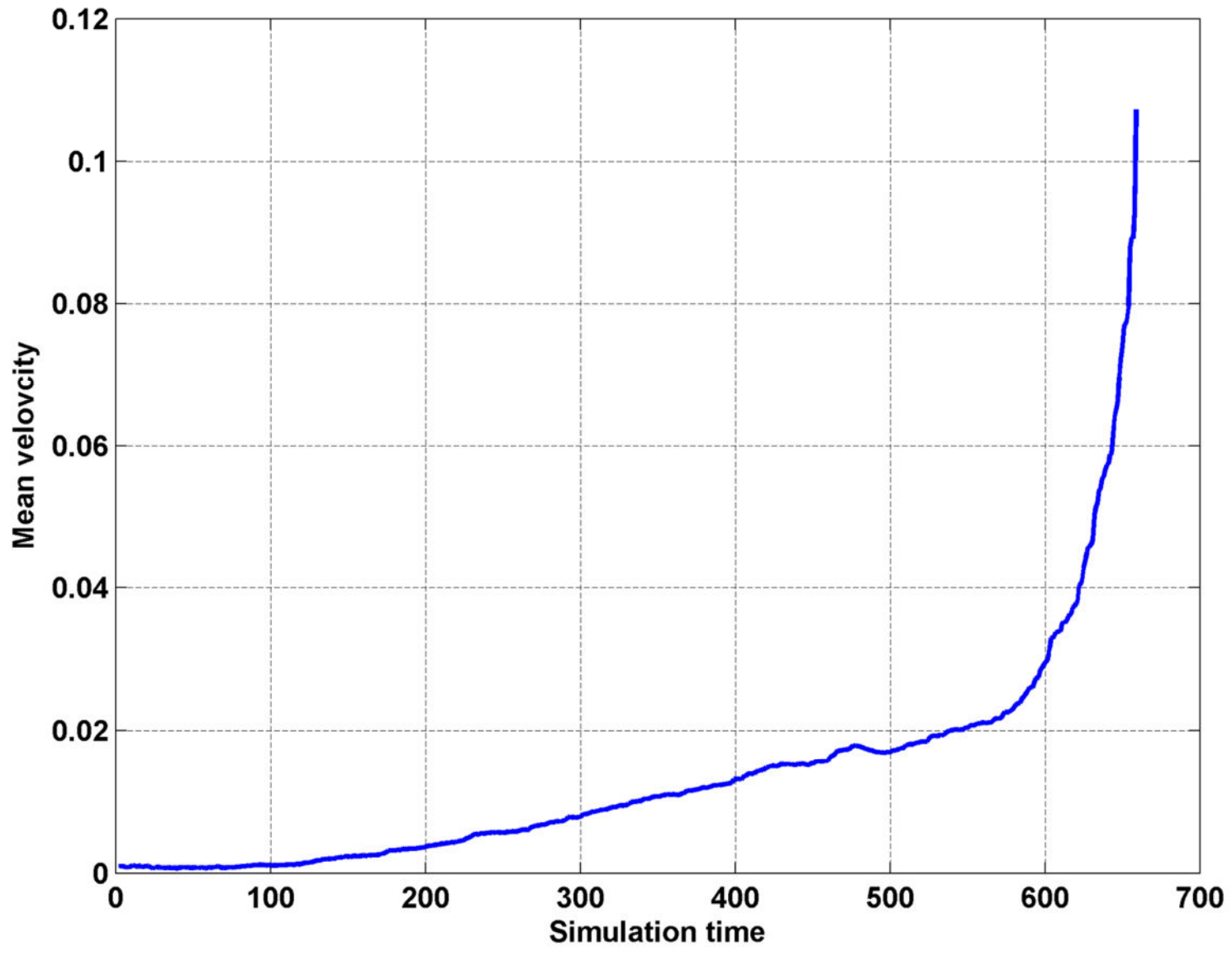

**Figure 9** Mean velocity versus simulation time

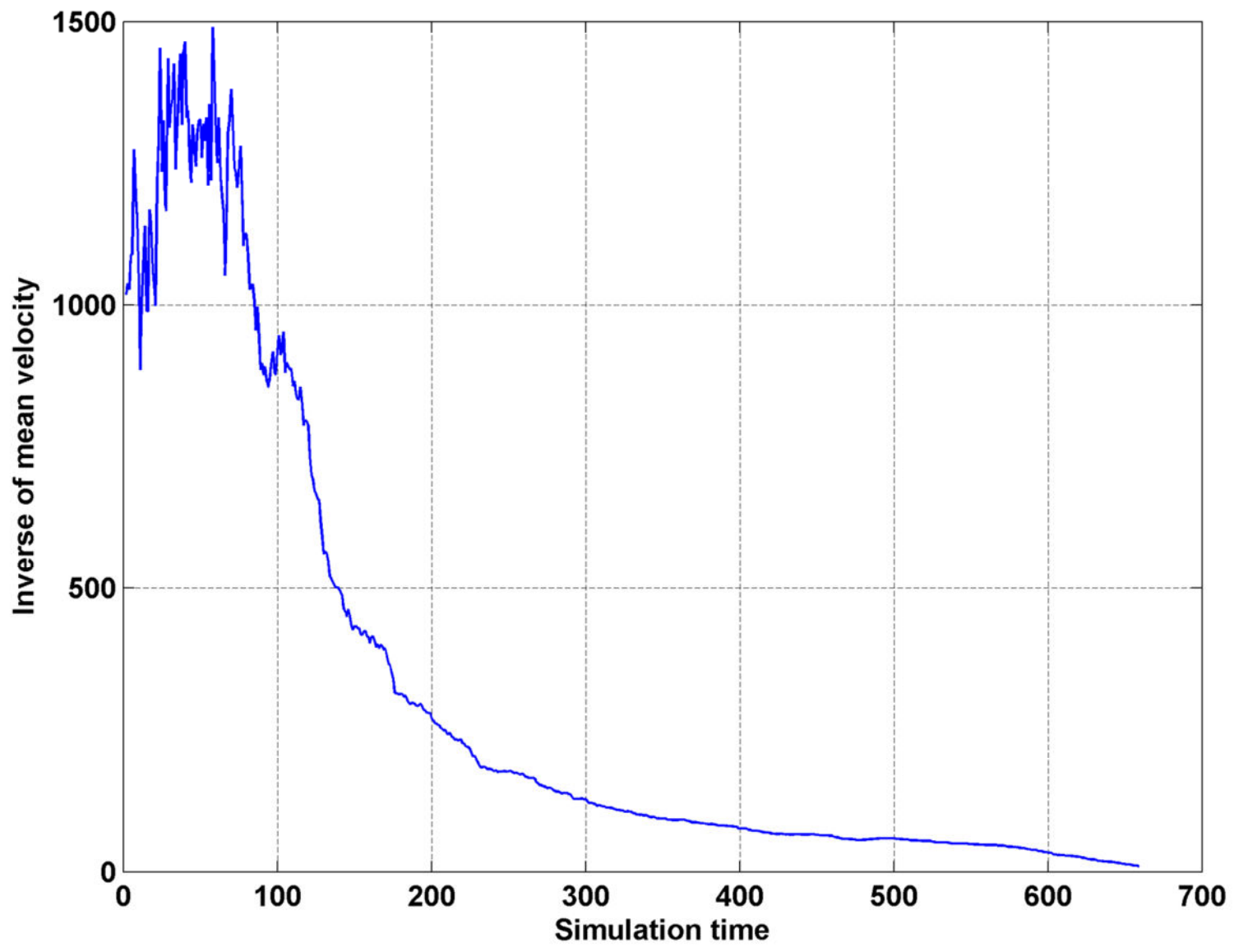

**Figure 10** Inverse of mean velocity versus simulation time

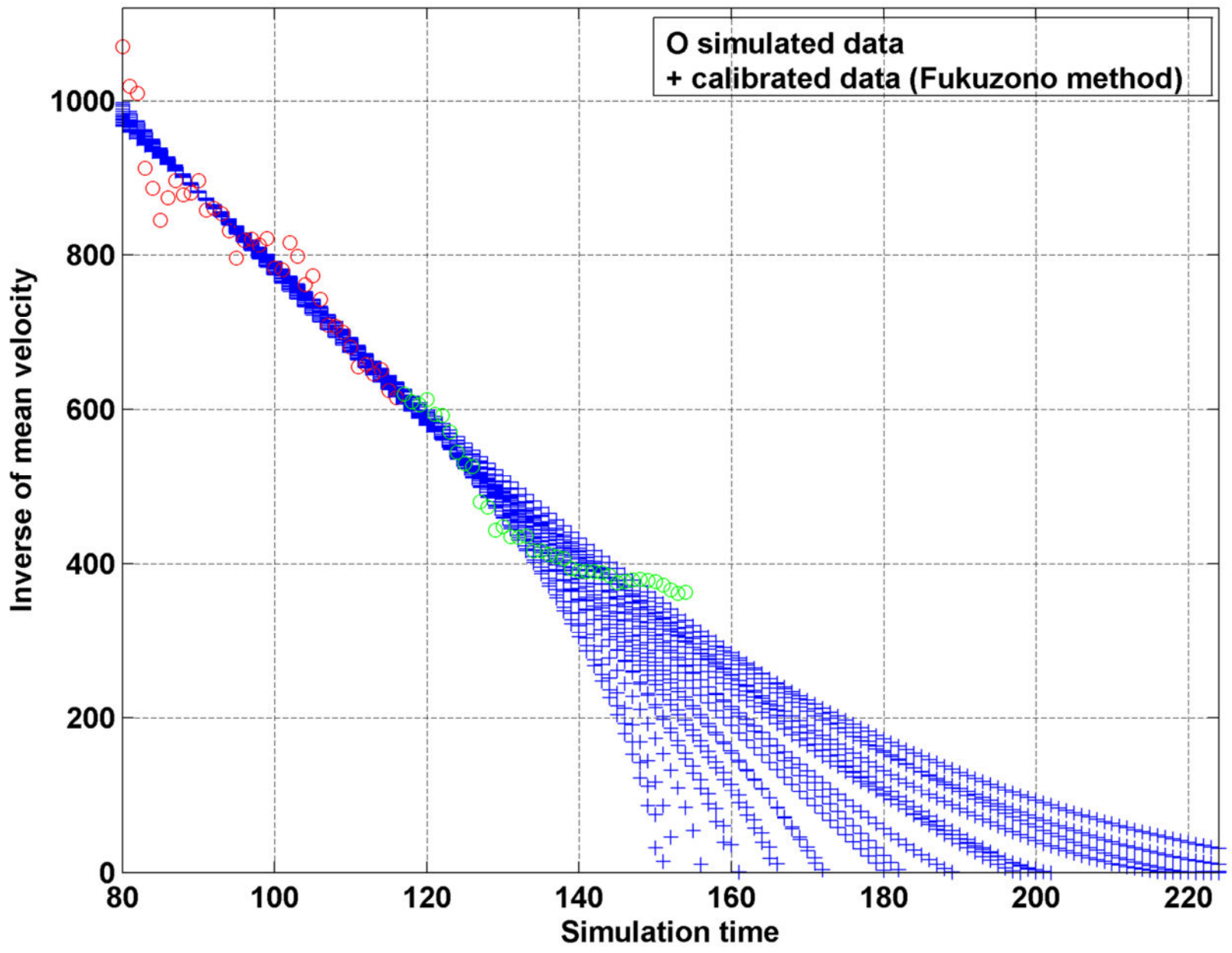

**Figure 11** Application of the inverse velocity method (Fukuzono 1985) to our simulated system

In Figs. 12(a) and 13(a) the landslide configuration is reported in the coordinate system of the slope ($x$-$z$) for the extreme values of the evaluated range of time triggering. We observe an initial motion of the upper horizontal layer and an initial phase of creep (in the system representation the green particles are in motion, while the red ones are at rest). In Figs. 12(b) and 13(b), the infiltration states for each position of slope are reported for $t$ = 150 and $t$ = 220. Moreover, in Fig. 14 we report a system configuration of the same simulation at $t$ = 600 where we note a slip phase with creep, detachments and arching phenomena.

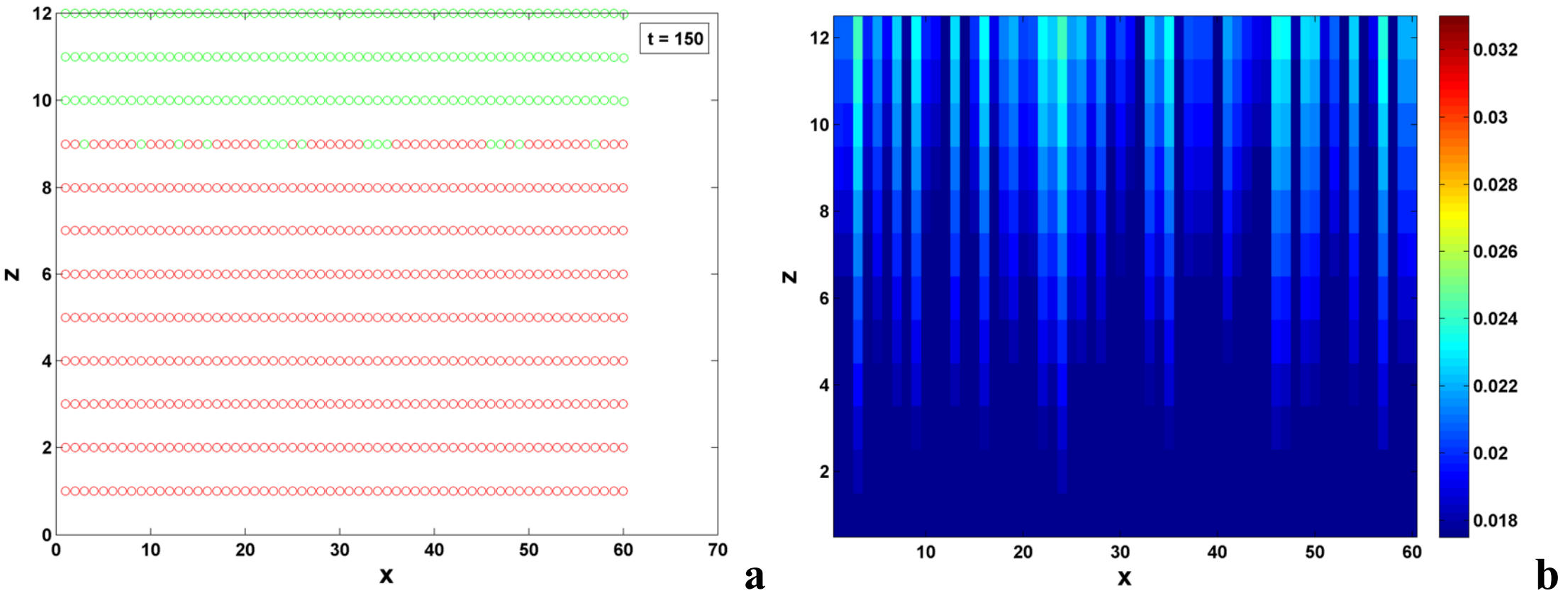

**Figure 12** **(a)** The simulated landslide in the coordinate system of the slope for $t$ = 150 **(b)** The simulated infiltration along the slope for $t$ = 150

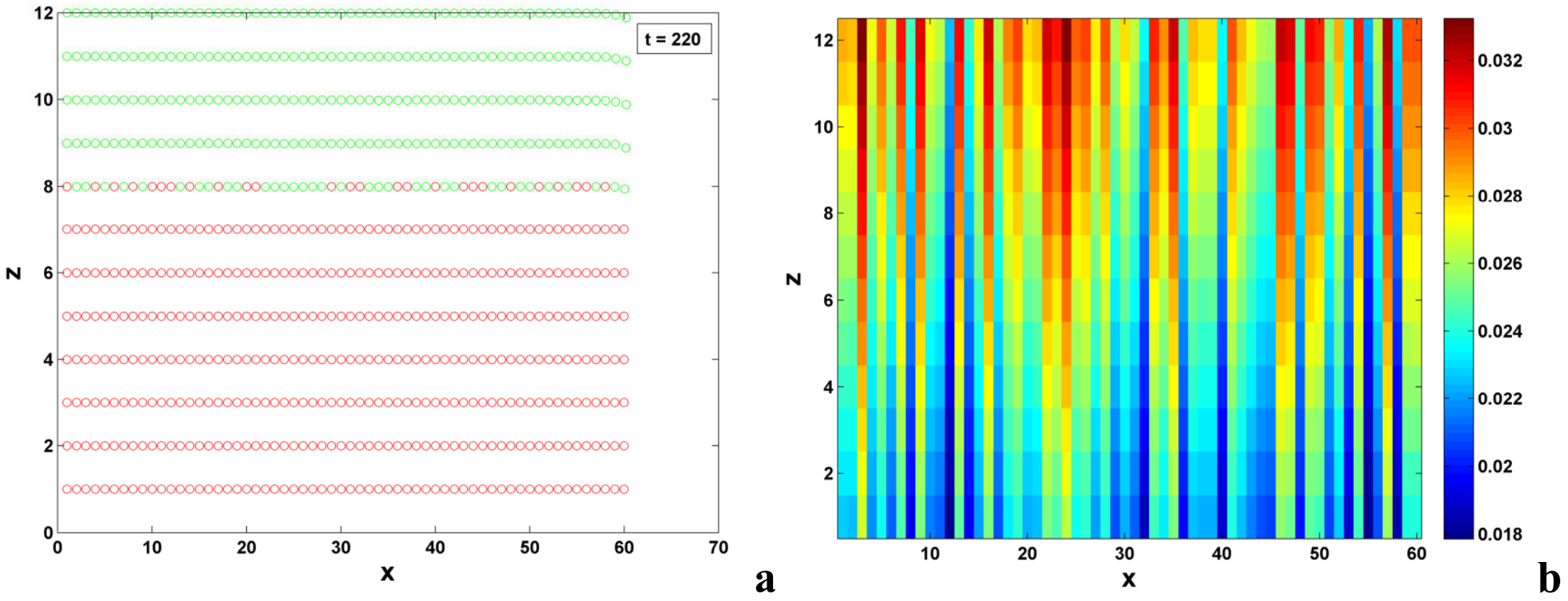

**Figure 13** **(a)** The simulated landslide in the coordinate system of the slope for $t$ = 220 **(b)** The simulated infiltration along the slope for $t$ = 220

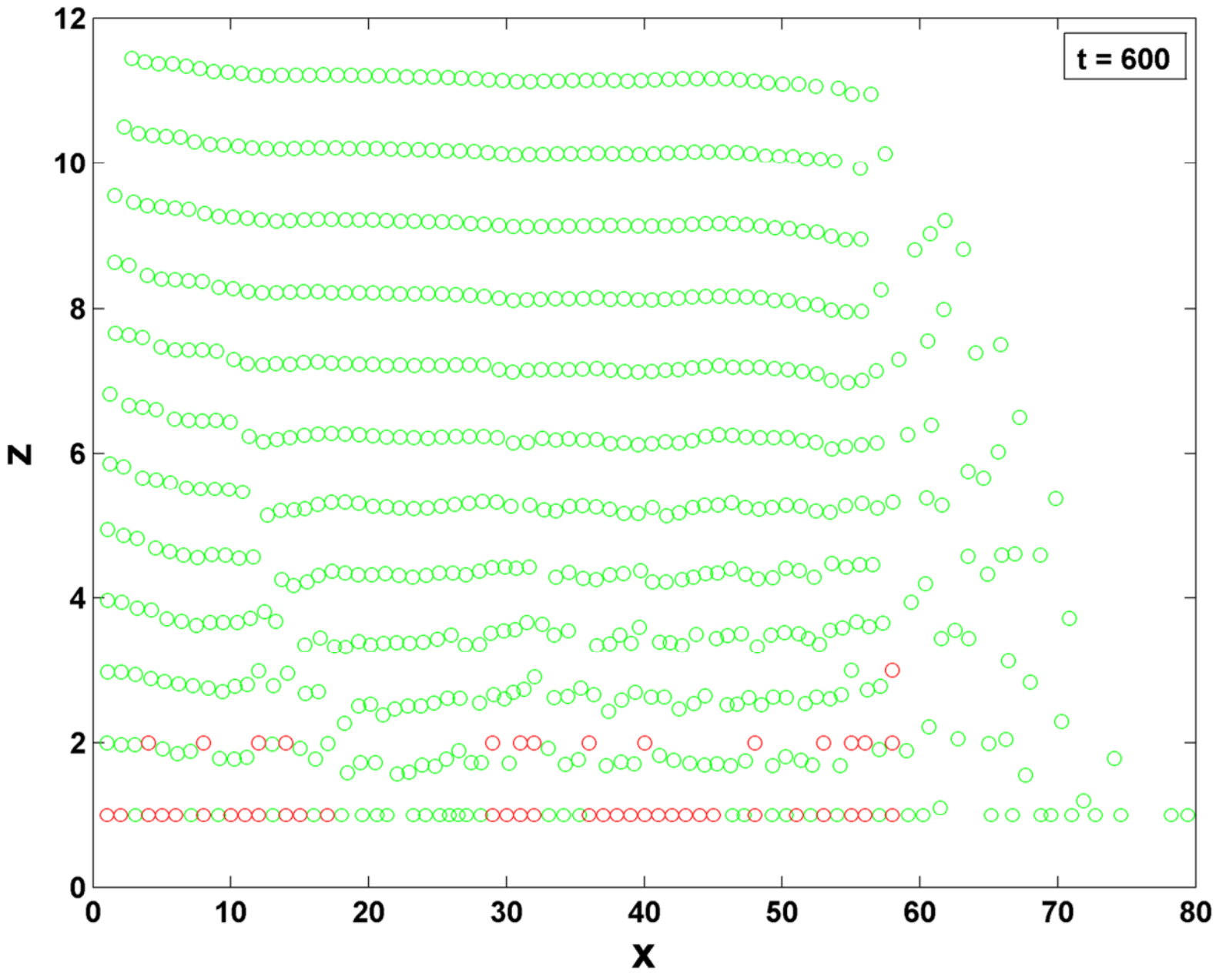

**Figure 14** The simulated landslide in the coordinate system of the slope for $t$ = 600

Other interesting results can be observed from a statistical point of view (Martelloni et al., 2012): we perform some simulations varying the viscosity coefficient $\mu$ of Eq. 15. We observe a transition of the mean energy increment distribution from Gaussian to power law after decreasing the viscosity coefficient from a finite initial value up to zero, as shown in Figs. 15, 16 and 17. This behavior is compatible with the corresponding velocity increasing of the landslide after decreasing the viscosity. In other words, this behavior is congruent with the stick-and-slip dynamics. Thus, the transition of the mean energy increment distribution is also observed in the same simulation at different times, i.e., by calculating this distribution in stick phase we observe a Gaussian distribution and not a power law even for a viscosity coefficient $\mu$ = 0, while considering the distribution in the slip phase we observe a power law also for high viscosity. Finally, we measured the time interval between successive time steps of the simulation ($t$, $t+1$) for which the masses start to move, i.e., we observe the distribution of the subsequent local triggering. In all simulations a power law distribution is observed (see in Fig. 18 the obtained result for $\mu = 0$). Finally in Tables 1 and 2 we reported the adopted fit estimators (Eqs. 21) and the optimal fit parameters ($a_1$, $b_1$, $c_1$, $a$, $b$) of the obtained distributions according to Eqs. 18, 19 and 20, i.e., Gaussian, log-normal and power law, respectively:

$$f(x) = a_1 \cdot \exp\left( -\left( \frac{x - b_1}{c_1} \right)^2 \right) \quad (18)$$

$$f(x) = a \cdot \exp(b) \quad (19)$$

$$f(x) = a \cdot x^b \quad (20)$$

where $x$ is the analyzed data.

The adopted estimators of the fitting accuracy are,

$$SSE = \sum_{i=1}^{n} \left(y_i - \hat{y}_i\right)^2$$

$$R^2 = 1 - \frac{SSE}{SST}; \quad SST = \sum_{i=1}^{n} \left(y_i - \bar{y}_i\right)^2$$

$$\hat{R}^2 = 1 - \left(1 - R^2\right)\frac{n-1}{n-p-1} \tag{21}$$

$$RMSE = \sqrt{\frac{SSE}{n-m}}$$

i.e., *SSE* is the *Sum of Squared Residuals*, $R^2$ is the *Coefficient of Determination*, $\hat{R}^2$ is *R Bar Squared* and *RMSE* is the *Root Mean Square Error*.

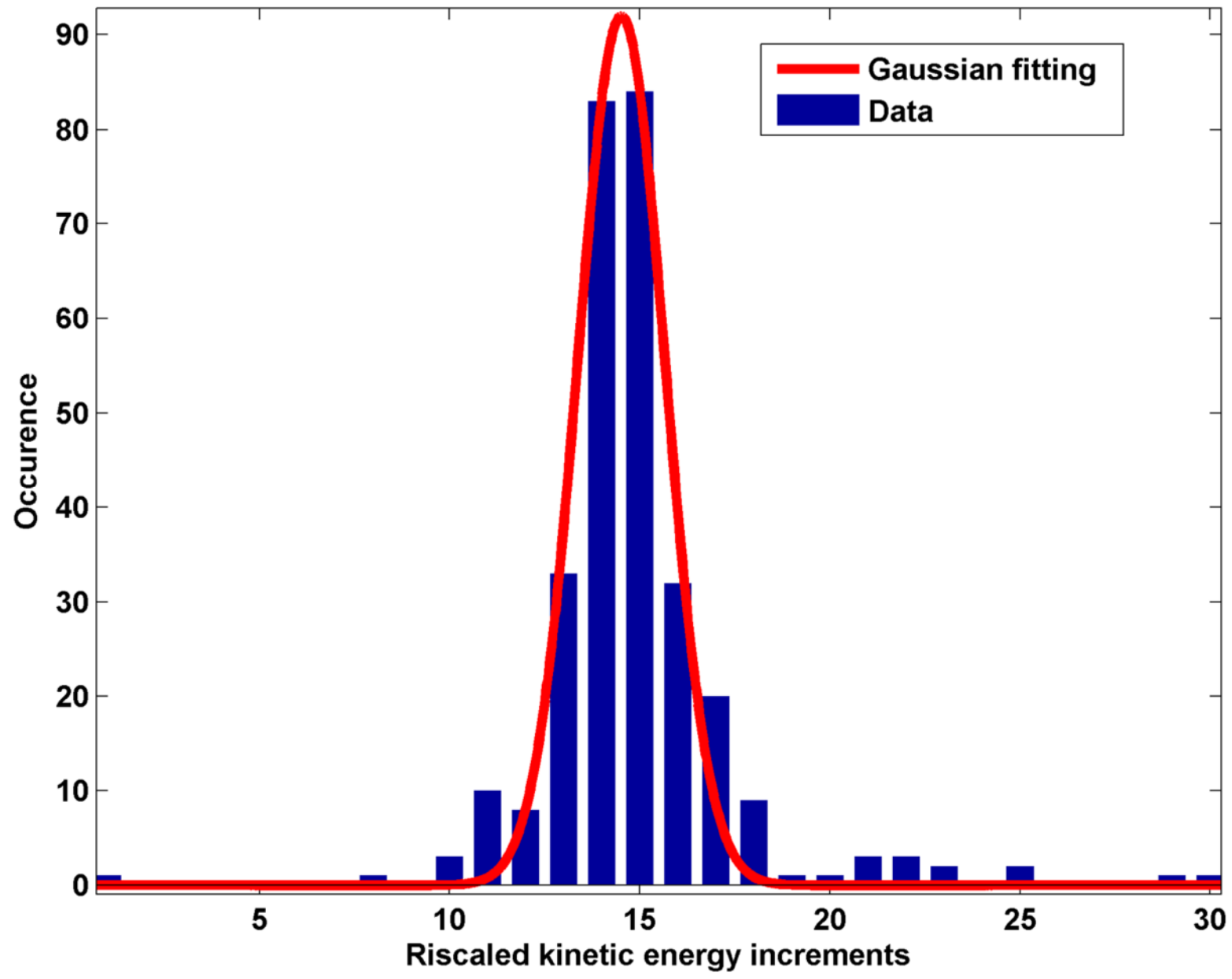

**Figure 15** The distribution of kinetic energy increments for $\mu = 0.01$

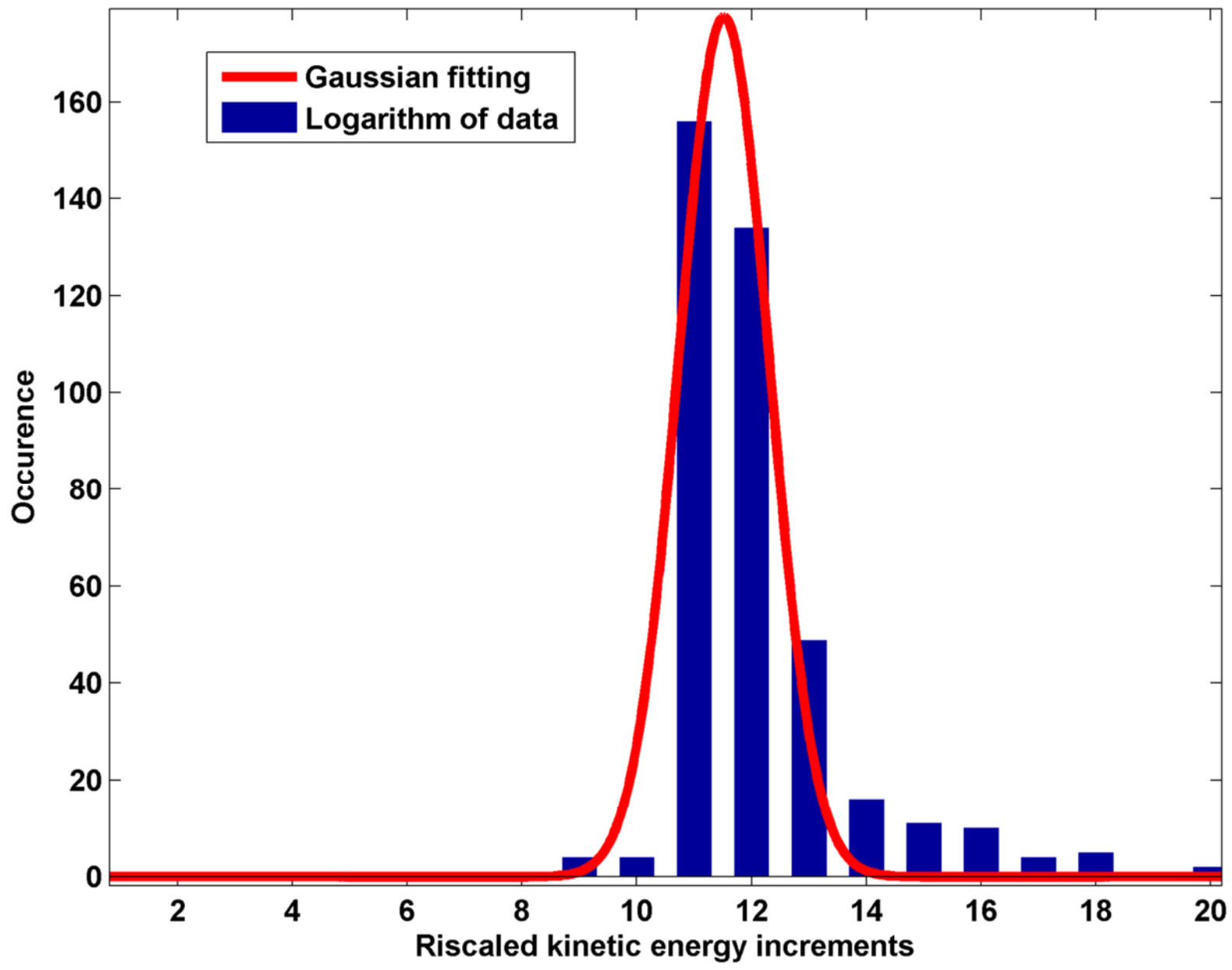

**Figure 16** The distribution of kinetic energy increments for $\mu = 0.0025$

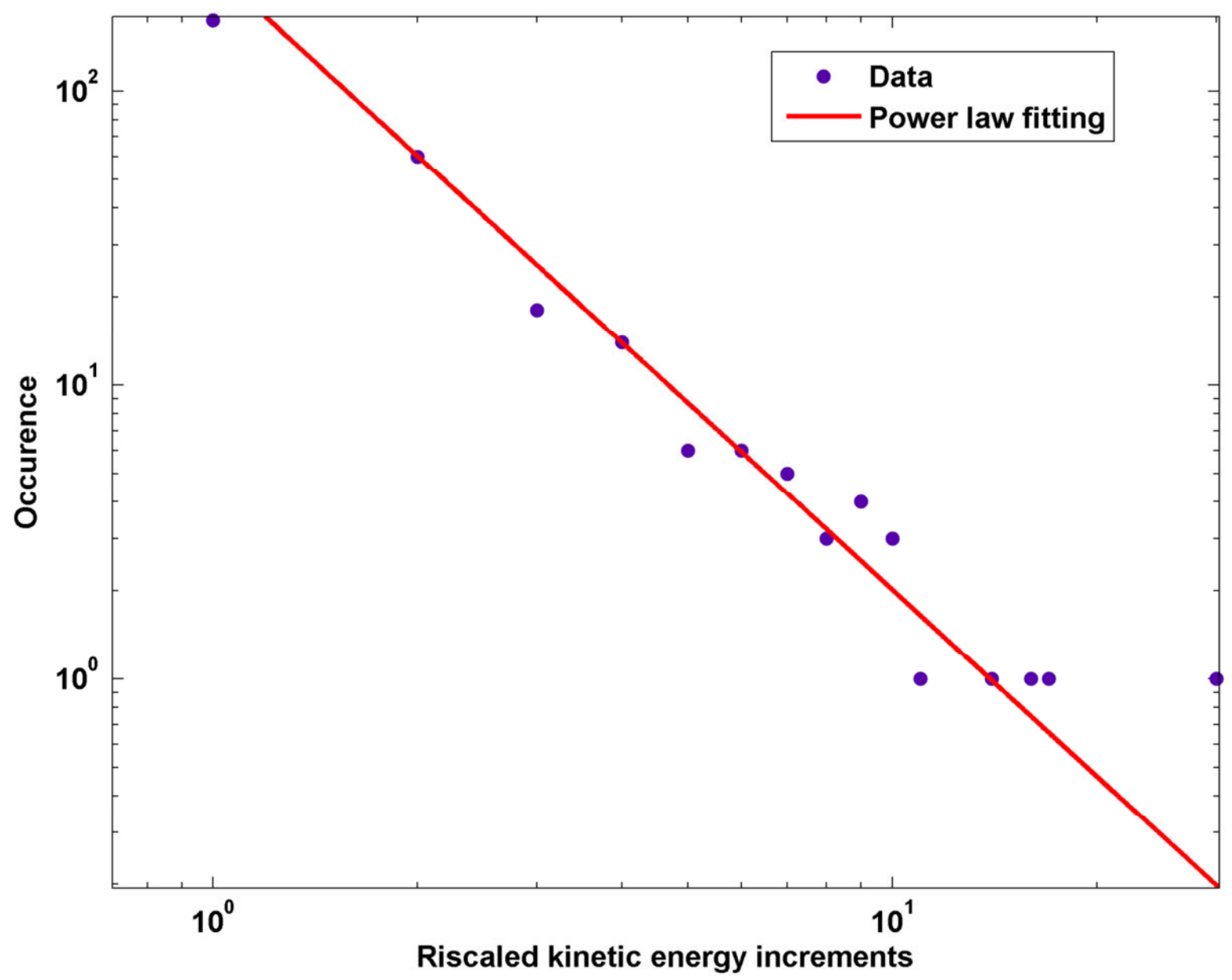

**Figure 17** The distribution of kinetic energy increments for $\mu$ = 0

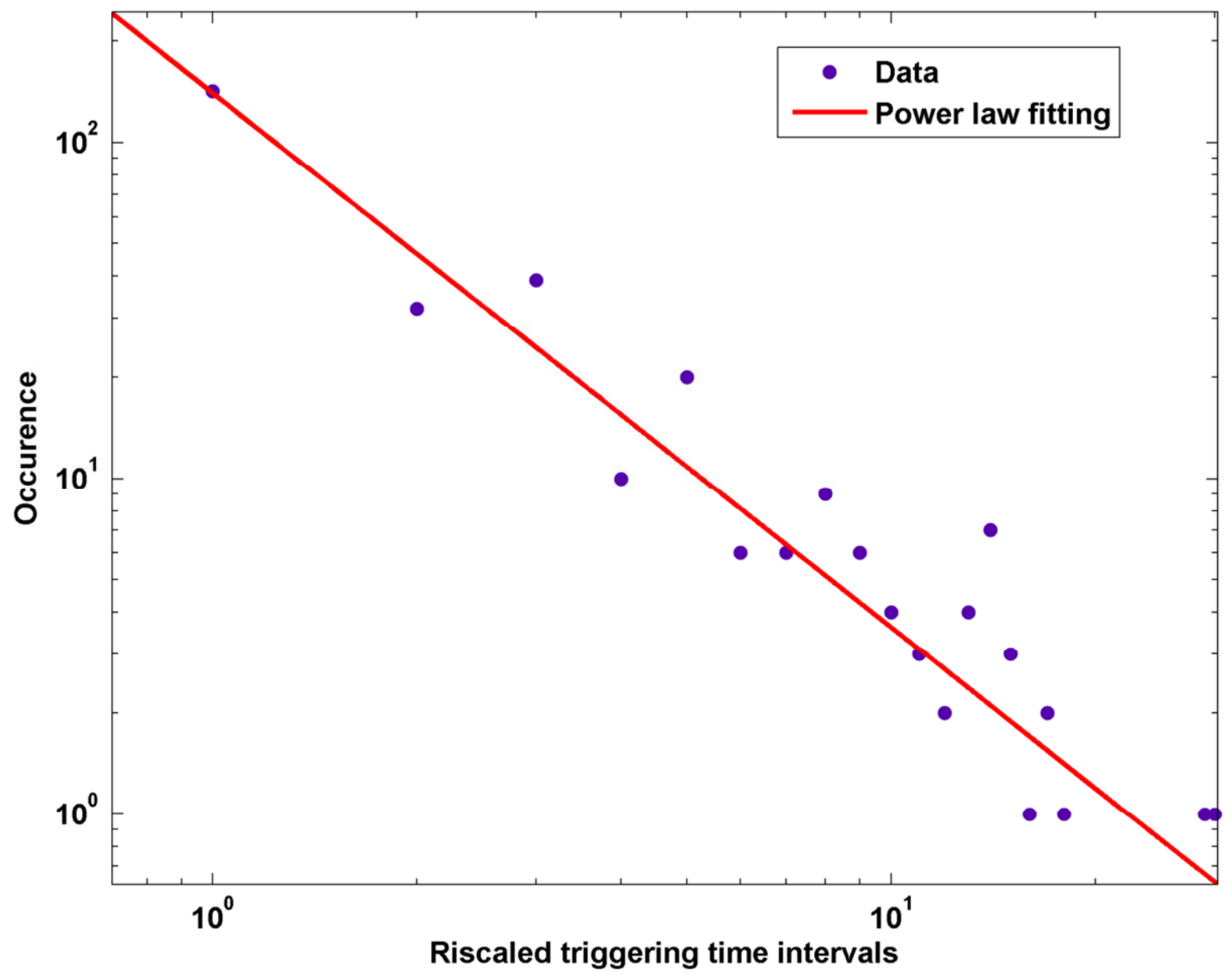

**Figure 18** The distribution of triggering time intervals for $\mu$ = 0

**Table 1** Kinetic energy increment distribution varying the coefficient of viscosity $\mu$, parameters of fit goodness and parameters of obtained distribution

| μ (Distribution) | 0.01 (Gaussian) | 0.0025 (Log-Normal) | 0 (Power law) | |
|---|---|---|---|---|
| SSE | 131.9 | 1752 | | |
| R-Square | 0.9904 | 0.9531 | SSE | 32.52 |
| Adjusted-R-Square | 0.9897 | 0.9475 | R-Square | 0.999 |
| RMSE | 2.21 | 10.15 | Adjusted-R-Square | 0.9989 |
| $a_1$ | 91.94 | 177.4 | RMSE | 1.078 |
| $b_1$ | 14.52 | 11.52 | a | 259.1 |
| $c_1$ | 1.612 | 1.108 | b | -2.11 |

**Table 2** Local time triggering distribution varying the coefficient of viscosity $\mu$, parameters of fit goodness and parameters of obtained power law distribution

| μ (Distribution) | 0.01 (Power law) | 0.0025 (Power law) | 0 (Power law) |
|---|---|---|---|
| SSE | 238.3 | 206.1 | 602.7 |
| R-Square | 0.9973 | 0.9936 | 0.9706 |
| Adjusted-R-Square | 0.9972 | 0.9933 | 0.9696 |
| RMSE | 2.917 | 2.713 | 4.639 |
| a | 353.1 | 174.4 | 139.9 |
| b | -1.722 | -1.469 | -1.59 |

# 4. Discussion and conclusions

In our opinion though the model proposed in this paper is still quite schematic, our results encourage for the researche in this direction. The results are consistent with the behavior of real landslides induced by rainfall and an interesting behavior emerges from the dynamic and statistical points of view. Emerging phenomena such as fractures, detachments and arching can be observed. In particular, the model reproduces well the energy and time distribution of avalanches, analogous to the observed Gutenberg-Richter and Omori power law distributions for earthquakes (Gutenberg and Richter 1956; Omori 1895). We note that other natural hazards (landslides, earthquakes and forest fires) also exhibit a power law distribution (Malamud et al., 2004; Turcotte 1997), characteristic of self-organized critical systems (Turcotte and Malamud 2004). Moreover, we observed an interesting statistical characteristic of this type of systems, i.e., a transition of the mean energy increment distribution from Gaussian to power law after decreasing the viscosity coefficient up to zero. This behavior is compatible with the corresponding velocity increase. The main advantage of these Lagrangian methods consists in the capability of following the trajectory of a single particle, possibly identifying its dynamical properties. Actually, we observed a characteristic velocity and energy pattern typical of a stick-and-slip dynamics, similar to real landslides behavior (Sornette et al., 2004). Moreover, we have shown that it is possible to apply the method of the inverse surface displacement velocity for predicting the failure time (Fukuzono 1985).

## Acknowledgements

We thank the Ente Cassa di Risparmio di Firenze for its support under the contract *Studio dei fenomeni di innesco e propagazione di frane in relazione ad eventi di pioggia e/o terremoti per mezzo di modelli matematici ed esperimenti di laboratorio su mezzi granulari*.